\documentclass[%
  reprint,
  showpacs,preprintnumbers,
  amsmath,amssymb,
  aps,
prb,
uplatex
]{revtex4-2}

\usepackage{graphicx}[]% Include figure files
\usepackage{dcolumn}% Align table columns on decimal point
\usepackage{bm}% bold math
\usepackage{multirow}
\usepackage{braket}
\usepackage{color}
\usepackage{ulem}

\usepackage{hyperref}
\usepackage{url}
\begin{document}

\preprint{APS/123-QED}

\title{
Rectangular and square skyrmion crystals on a centrosymmetric square lattice\\ with easy-axis anisotropy
}

\author{Satoru Hayami}
\affiliation{
Department of Applied Physics, The University of Tokyo, Tokyo 113-8656, Japan
}
 
\begin{abstract}
We report our theoretical results on the emergence of rectangular- and square-shaped magnetic skyrmion crystals on a centrosymmetric square lattice with magnetic anisotropy. 
By performing the simulated annealing for a frustrated spin model with an easy-axis single-ion anisotropy on a square lattice, we find that a competition of the interactions in momentum space gives rise to the formation of two skyrmion crystals, where the skyrmion cores are aligned in a rectangle- or square-lattice way.  
Especially, the interactions at the higher harmonic wave vectors play an important role in stabilizing the square/rectangular skyrmion crystals. 
Our results provide a reference to explore further skyrmion-hosting materials in centrosymmetric tetragonal systems. 
\end{abstract}
\maketitle

\section{Introduction}
Topological spin textures with a spin scalar chirality by a triple scalar product $\bm{S}_i \cdot (\bm{S}_j \times \bm{S}_k)$ have recently attracted great interest, since they lead to unusual quantum transports originating from the spin Berry phase~\cite{berry1984quantal,Loss_PhysRevB.45.13544,Ye_PhysRevLett.83.3737,Xiao_RevModPhys.82.1959}. 
Depending on spatial distribution of the scalar chirality or topological charge, a variety of topological spin textures, such as a skyrmion and meron on a two-dimensional plane and a hedgehog and hopfion in a three-dimensional space, are defined~\cite{nagaosa2013topological,batista2016frustration,Tokura_doi:10.1021/acs.chemrev.0c00297,hayami2021topological,gobel2021beyond}. 
As they are topologically-protected states and exhibit rich electromagnetic phenomena against external stimuli, such as a topological Hall effect~\cite{Bruno_PhysRevLett.93.096806,Neubauer_PhysRevLett.102.186602,Kanazawa_PhysRevLett.106.156603}, Nernst effect~\cite{Shiomi_PhysRevB.88.064409,mizuta2016large}, and nonreciprocal transport~\cite{Hamamoto_PhysRevB.95.224430,seki2020propagation}, it is expected to have a great advantageous in applications to next-generation spintronics devices~\cite{romming2013writing,fert2013skyrmions,fert2017magnetic,zhang2020skyrmion}. 

These topological spin textures often appear in the periodic form, such as a skyrmion crystal (SkX) and a hedgehog lattice, as observed in various magnetic materials~\cite{Muhlbauer_2009skyrmion,yu2010real,tanigaki2015real,fujishiro2019topological,Tokura_doi:10.1021/acs.chemrev.0c00297}. 
When the topological objects are aligned periodically, there is a degree of freedom in terms of packing. 
Although most of the SkXs observed in materials have been identified as a triangular-lattice SkX (T-SkX)~\cite{Muhlbauer_2009skyrmion,yu2010real}, a square-lattice SkX (S-SkX) has also been observed in the noncentrosymmetric magnets Co$_{10-x/2}$Zn$_{10-x/2}$Mn$_x$~\cite{tokunaga2015new,karube2016robust,karube2018disordered,Karube_PhysRevB.102.064408,henderson2021skyrmion} and Cu$_2$OSeO$_3$~\cite{chacon2018observation,takagi2020particle}, and the centrosymmetric magnet GdRu$_2$Si$_2$~\cite{khanh2020nanometric,Yasui2020imaging}. 

From the energetic viewpoint, the emergence of the T-SkX seems to be natural, since the T-SkX is constructed by a superposition of three spiral waves with the ordering vectors $\bm{Q}_1$, $\bm{Q}_2$, and $\bm{Q}_3$, where the angle among the constituent waves is 120$^{\circ}$ to satisfy $\bm{Q}_1+\bm{Q}_2+\bm{Q}_3=\bm{0}$. 
This relation indicates the appearance of an effective interaction in the form of $(\bm{S}_{\bm{0}}\cdot \bm{S}_{\bm{Q}_1})(\bm{S}_{\bm{Q}_2}\cdot \bm{S}_{\bm{Q}_3})$ in the Ginzburg-Landau free energy. 
Moreover, the above condition leads to the suppression of the energy loss by the exchange and/or Dzyaloshinskii-Moriya (DM) interaction~\cite{dzyaloshinsky1958thermodynamic,moriya1960anisotropic} in terms of the higher-harmonic contribution that arises from the superposition of multiple spirals~\cite{hayami2021field}. 

On the other hand, the S-SkX consisting of two spirals along the $\bm{Q}_1$ and $\bm{Q}_2$ directions, which are connected by $90^{\circ}$ rotation, is rather difficult to stabilize as a thermodynamic stable state only when considering the effect of the DM interaction~\cite{rossler2006spontaneous,Yi_PhysRevB.80.054416} and frustrated exchange interaction~\cite{Okubo_PhysRevLett.108.017206,leonov2015multiply,Lin_PhysRevB.93.064430,Hayami_PhysRevB.93.184413} in contrast to the T-SkX. 
Recently, theoretical studies have revealed that the S-SkX is stabilized by introducing the additional magnetic anisotropy and dipole-dipole interactions on the basis of the DM and frustrated interactions~\cite{Yi_PhysRevB.80.054416,Hayami_PhysRevLett.121.137202,Utesov_PhysRevB.103.064414,Wang_PhysRevB.103.104408}. 
It was also demonstrated that the S-SkX appears in the ground state by considering the multiple-spin interaction arising from the spin-charge coupling in centrosymmetric itinerant magnets~\cite{Hayami_PhysRevB.103.024439,Hayami_doi:10.7566/JPSJ.89.103702,Hayami_PhysRevB.105.104428}. 
Simultaneously, the recent experimental studies indicate the emergence of the topological spin textures in the centrosymmetric tetragonal system in EuAl$_4$~\cite{Shang_PhysRevB.103.L020405,kaneko2021charge,Zhu2022}, EuGa$_4$~\cite{zhang2021giant,Zhu2022}, EuGa$_2$Al$_2$~\cite{moya2021incommensurate}, and Mn$_{2-x}$Zn$_x$Sb~\cite{Nabi_PhysRevB.104.174419}, which might host the S-SkX. 
Thus, it is highly desired to clarify the essence of the stabilization of the S-SkX at the microscopic level.

In the present study, we investigate the instability toward the S-SkX by focusing on the frustration in momentum space. 
We show that the interplay between the competing interactions in momentum space and the easy-axis single-ion anisotropy gives rise to the S-SkX in an external magnetic field by the simulated annealing. 
Furthermore, we obtain another SkX, where the skyrmion core is extended in a rectangle way (R-SkX) when the frustration in momentum space is maximized. 
We also find a plethora of double-$Q$ (2$Q$) states with the local scalar chirality but without the global one. 
Our results provide a reference to not only understand the origin of the SkX but also to identify magnetic phases observed in experiments.

The rest of the paper is organized as follows. 
In Sec.~\ref{sec:Square skyrmion crystal}, we introduce the double-$Q$ S-SkX on the square lattice. 
We also discuss the competing exchange interactions in momentum space, which is the origin of the S-SkX and R-SkX. 
In Sec.~\ref{sec:Model}, we present the spin model including the momentum-resolved interactions and the easy-axis single-ion anisotropy. 
We discuss the instabilities toward the S-SkX and R-SkX by performing the simulated annealing.  
Section~\ref{sec:Summary} is devoted to a summary.

\section{Square skyrmion crystal}
\label{sec:Square skyrmion crystal}
\begin{figure}[t!]
\begin{center}
\includegraphics[width=0.75 \hsize ]{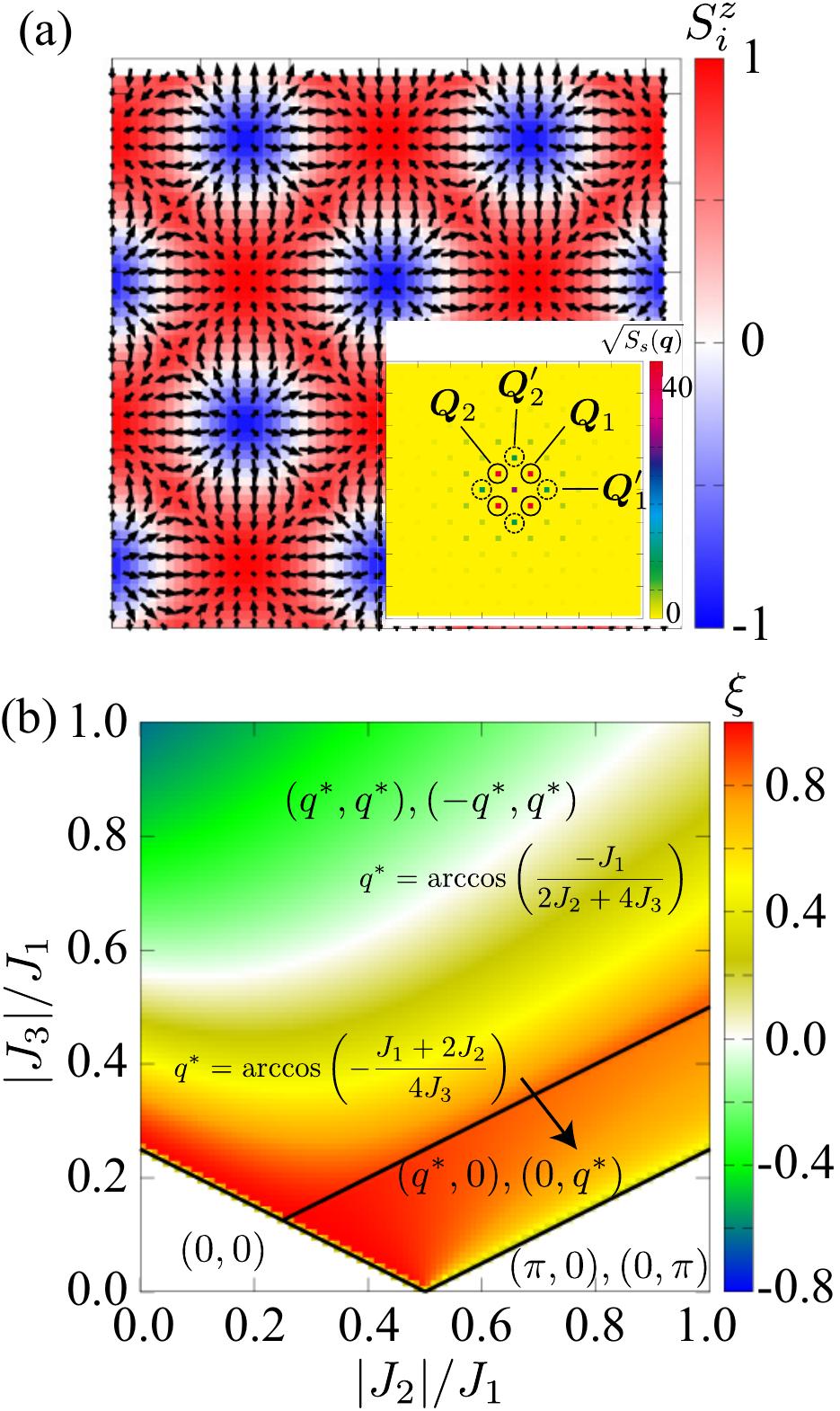} 
\caption{
\label{fig:fig1}
(a) Real-space spin configuration of the S-SkX in Eq.~(\ref{eq:SkX}). 
The arrows and the contour show $(S_i^x, S_i^y)$ and $S_i^z$, respectively. 
The inset of (a) represents the spin structure factor in the first Brillouin zone. 
The solid (dashed) circles represent $\bm{Q}_1$ and $\bm{Q}_2$ ($\bm{Q}'_1=\bm{Q}_1-\bm{Q}_2$ and $\bm{Q}'_2=\bm{Q}_1+\bm{Q}_2$). 
(b) Contour plot of $\xi$ in the plane of $J_2$ and $J_3$. 
The optimal ordering vectors are also shown in each region. 
In the region for the ordering vectors $\bm{Q}_1=(q^*,q^*)$ and $\bm{Q}_2=(-q^*,q^*)$ [$\bm{Q}'_1=(q^*,0)$ and $\bm{Q}'_2=(0,q^*)$], $\xi=J_{\bm{Q}_1+\bm{Q}_2}/J_{\bm{Q}_1}$ ($\xi=J_{(\bm{Q}'_1+\bm{Q}'_2)/2}/J_{\bm{Q}'_1}$).  
}
\end{center}
\end{figure}

Let us start by presenting the spin texture in the S-SkX, which is described by a superposition of two spiral waves as
\begin{align}
\label{eq:SkX}
\bm{S}_i \propto 
\left(
    \begin{array}{c}
   - \cos \mathcal{Q}_1 +  \cos \mathcal{Q}_2 \\
   - \cos \mathcal{Q}_1 -  \cos \mathcal{Q}_2 \\
   - a_z (\sin \mathcal{Q}_1+\sin \mathcal{Q}_2)+ \tilde{M}_z
          \end{array}
  \right)^{\rm T}, 
\end{align}
where $\mathcal{Q}_\eta=\bm{Q}_\eta \cdot \bm{r}_i+\theta_\eta$ for $\eta=1,2$ and $\bm{Q}_1 \perp \bm{Q}_2$; $a_z$ and $\tilde{M}_z$ depend on the model parameters and $|\bm{S}_i|=1$. 
The spin texture with $\bm{Q}_1=(\pi/8,\pi/8)$ and $\bm{Q}_2=(-\pi/8,\pi/8)$ in Eq.~(\ref{eq:SkX}) is shown in Fig.~\ref{fig:fig1}(a), where the skyrmion cores are located at the center of the square plaquette by taking $\theta_1=\pi/8$ and $\theta_2=0$~\cite{Hayami_PhysRevResearch.3.043158,hayami2021phase} and form the square lattice. 
This spin configuration exhibits a quantized skyrmion number of $-1$. 
The corresponding spin structure factor $S_s(\bm{q})=\sum_{\alpha}S^{\alpha}_s(\bm{q})=(1/N) \sum_{ij} S^{\alpha}_i S^{\alpha}_j e^{i \bm{q}\cdot (\bm{r}_i-\bm{r}_j)}$ ($N$ is the system size and $\alpha=x,y,z$) is shown in the inset of Fig.~\ref{fig:fig1}(a); the maximum appears at $\bm{Q}_1$ and $\bm{Q}_2$ denoted as the solid circles, while the second maximum except for $\bm{q}=\bm{0}$ appears at $\bm{Q}_1-\bm{Q}_2 \equiv \bm{Q}'_1$ and $\bm{Q}_1+\bm{Q}_2 \equiv \bm{Q}'_2$ denoted as the dashed circles. 
The appearance of the peaks at $\bm{Q}'_1$ and $\bm{Q}'_2$ is owing to a superposition of the spirals with $\bm{Q}_1$ and $\bm{Q}_2$ as found in the $z$-spin component in Eq.~(\ref{eq:SkX}), which corresponds to the higher harmonics.  
This indicates that the interactions for the higher-harmonic wave vectors ($\bm{Q}'_1$ and $\bm{Q}'_2$) might be important to determine the optimal spin configuration when the interaction at $\bm{Q}'_1$ and $\bm{Q}'_2$ is comparable to that at $\bm{Q}_1$ and $\bm{Q}_2$. 

\begin{figure*}[t!]
\begin{center}
\includegraphics[width=1.0 \hsize ]{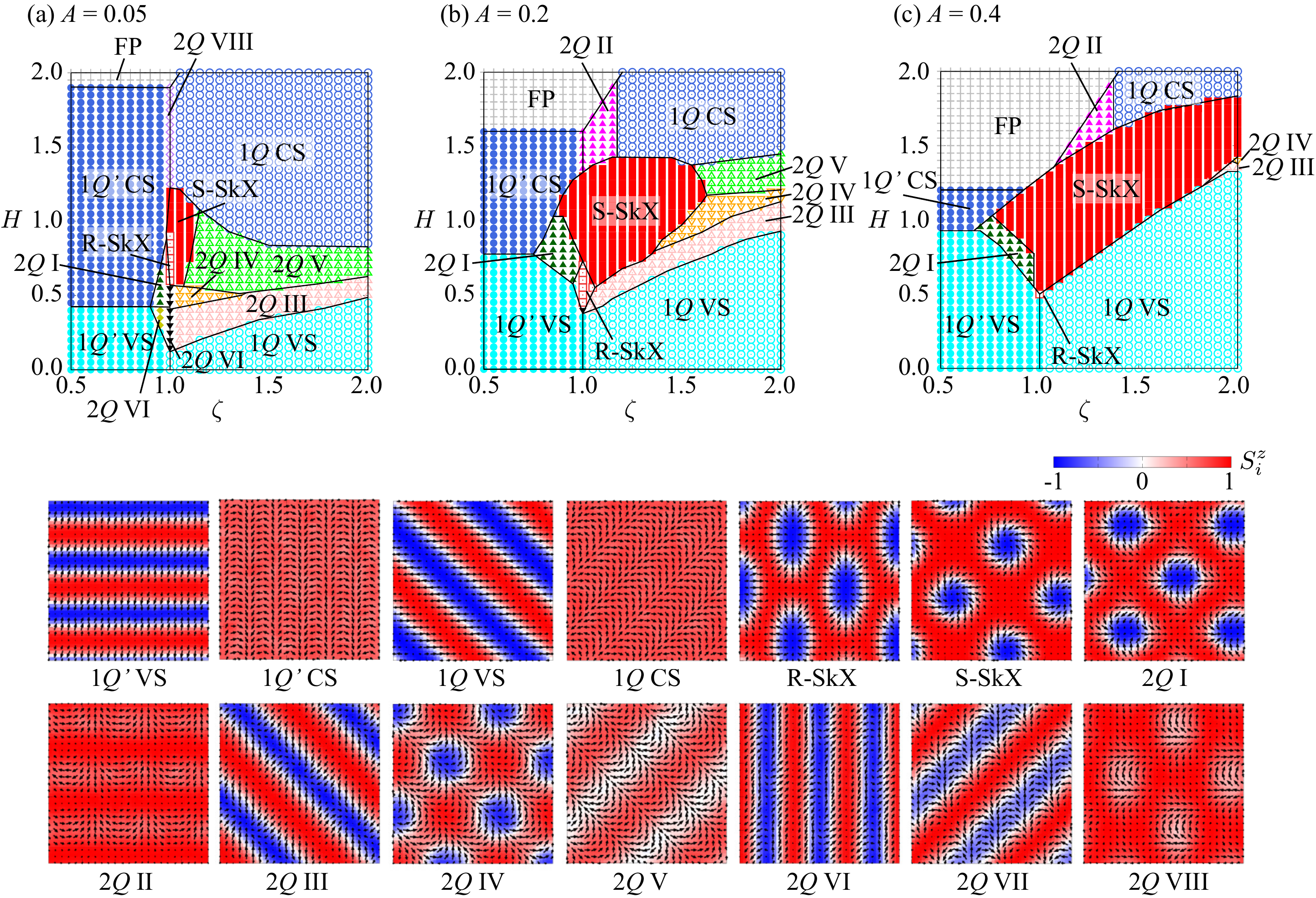} 
\caption{
\label{fig:fig2}
Phase diagrams while varying $\zeta=J/J'$ and magnetic field $H$ at (a) $A=0.05$, (b), $A=0.2$, and $A=0.4$ obtained by the simulated annealing for $N=32^2$. 
S-SkX, R-SkX, VS, CS, and FP stand for the square skyrmion crystal, rectangular skyrmion crystal, vertical spiral, conical spiral, and fully-polarized phases, respectively. 
The lower panel shows the spin configurations, where the arrows and the contour show $(S_i^x, S_i^y)$ and $S_i^z$, respectively; see also Table~\ref{table:mp} for the detailed spin textures in each phase. 
}
\end{center}
\end{figure*}

Such a situation can happen under the competing exchange interactions in real space. 
We consider the Heisenberg interactions up to the third neighbors, $J_1$, $J_2$, and $J_3$, on the square lattice, where the lattice constant is set to be unity.  
Then, the interaction in momentum space $J_{\bm{q}}$ in $-\sum_{\bm{q}}J_{\bm{q}}\bm{S}_{\bm{q}}\cdot \bm{S}_{-\bm{q}}$ is given by 
\begin{align}
\label{eq:Jq}
J_{\bm{q}}&= 2J_1 (c_x + c_y)+4J_2 c_{x}c_{y}+2J_3 (c_{2x} + c_{2y}), 
\end{align}
where $c_{\mu}=\cos q_{\mu}$ and $\mu=x,y$. 
By considering the (anti)ferromagnetic interaction for $J_1>0$ ($J_2, J_3<0$), the optimal ordering vectors in the plane of $J_2$ and $J_3$ are shown in Fig.~\ref{fig:fig1}(b)~\cite{Wang_PhysRevB.103.104408}. 
The color map shows the ratio of the harmonic contribution to $J_{\bm{q}}$ normalized by the largest value, $\xi$; $\xi=J_{\bm{Q}_1+\bm{Q}_2}/J_{\bm{Q}_1}$ ($\xi=J_{(\bm{Q}'_1+\bm{Q}'_2)/2}/J_{\bm{Q}'_1}$), where $\bm{Q}_\eta$ ($\bm{Q}'_\eta$) is the ordering wave vector along the $\langle 110 \rangle$ ($\langle 100 \rangle$) direction, which gives the largest $J_{\bm{q}}$. 
In the vicinity of the phase boundaries among the states with $\bm{q}=(0,0)$, $(q^*,0)$, and $(q^*, q^*)$, the interactions at ${\bm{Q}_1}$ and ${\bm{Q}_1+\bm{Q}_2}$ become comparable, e.g., $J_{\bm{Q}_1+\bm{Q}_2}/J_{\bm{Q}_1}\simeq 0.94$ for $J_2=-0.35$ and $J_3=-0.18$. 

\section{Model}
\label{sec:Model}
We focus on the instability toward the SkXs in the above situation. 
For that purpose, we consider a simplified spin Hamiltonian as 
\begin{align}
\label{eq:Ham}
\mathcal{H}= -\sum_{\eta} J_{\bm{Q}_\eta} \bm{S}_{\bm{Q}_{\eta}} \cdot \bm{S}_{-\bm{Q}_{\eta}}
-A\sum_{i} (S_i^z)^2 - H \sum_i S_i^z, 
\end{align}
where we take into account the dominant interactions with eight $\bm{Q}_\eta$; $\pm \bm{Q}_1=(Q,Q)$, $\pm \bm{Q}_2=(-Q,Q)$, $\pm \bm{Q}_3\equiv\bm{Q}'_1=(2Q,0)$, and $\pm  \bm{Q}_4\equiv \bm{Q}'_2=(0,2Q)$ in order to discuss the instability toward the S-SkX and other multiple-$Q$ spin configurations in an efficient way~\cite{leonov2015multiply,Hayami_PhysRevB.103.224418,hayami2021skyrmion,hayami2022skyrmion}. 
Meanwhile, we neglect the instability toward the T-SkX by implicitly supposing the fourfold magnetic anisotropy that arises from the tetragonal geometry, although we do not explicitly consider such an effect for simplicity~\cite{Wang_PhysRevB.103.104408}. 
The second and third terms in Eq.~(\ref{eq:Ham}) stand for the easy-axis single-ion anisotropy with $A>0$ and the Zeeman coupling to an external magnetic field with $H>0$. 
In the following, we set $J_{\bm{Q}'_1}=J_{\bm{Q}'_2}\equiv J'$ and  $J_{\bm{Q}_1}=J_{\bm{Q}_2}\equiv J $; $J'$ is taken as the energy unit ($J'=1$). 
Although $J$ and $J'$ in Eq.~(\ref{eq:Ham}) are expected to be determined by the frustrated exchange interaction in Eq.~(\ref{eq:Jq}), we treat $J$ as a phenomenological parameter so as to cover the situation with the different $J_{\bm{q}}$ beyond the third-neighbor interactions. 
Accordingly, we fix $Q=\pi/8$, although the following results hold for other finite-$Q$ ordering vectors. 
It is noted that a similar situation can be realized by considering the long-ranged Ruderman-Kittel-Kasuya-Yosida interaction when the bare susceptibility shows the maxima at finite-$Q$ ordering wave vectors, as studied in tetragonal~\cite{Ozawa_doi:10.7566/JPSJ.85.103703,Hayami_PhysRevB.95.224424,Hayami_PhysRevB.103.024439}, hexagonal~\cite{Ozawa_PhysRevLett.118.147205,Hayami_PhysRevB.95.224424,Wang_PhysRevLett.124.207201,Hayami_PhysRevB.103.054422}, and trigonal systems~\cite{yambe2021skyrmion}.

\section{Result}
\label{sec:Result}
Figures~\ref{fig:fig2}(a)-\ref{fig:fig2}(c) show the low-temperature phase diagrams at (a) $A=0.05$, (b) $A=0.2$, and (c) $A=0.4$ while varying $\zeta=J/J'$ and $H$ by performing the simulated annealing following the manner in Refs.~\cite{Hayami_PhysRevB.95.224424,hayami2020multiple}, where the final temperature and the system size are taken at $T=0.01$ in the unit of $J'$ and $N=32^2$, respectively. 
We show the real-space spin configurations for fourteen magnetic phases, which are obtained by the simulations in the lower panel of Fig.~\ref{fig:fig2}. 
In addition, we list nonzero magnetic moments with $\bm{Q}_\eta$ and $\bm{Q}'_\eta$ given by $m^z_{\bm{q}}=\sqrt{S^z_{s}(\bm{q})/N}$ and $m^{xy}_{\bm{q}}=\sqrt{\{S^x_{s}(\bm{q})+S^y_{s}(\bm{q})\}/N}$ in Table~\ref{table:mp}. 

When the effect of the frustration in momentum space is small, i.e., $\zeta \gg 1$ or $\zeta \ll 1$, the single-$Q$ (1$Q$) vertical spiral (VS) and cycloidal spiral (CS) states are stabilized in the low- and high-field regions for small $A$, respectively; the ordering vectors lie at $\bm{Q}_1$ or $\bm{Q}_2$ for $\zeta>1$, while those are at $\bm{Q}'_1$ or $\bm{Q}'_2$ for $\zeta<1$. 

The instability toward the S-SkX occurs around $\zeta \simeq 1$, where the effect of the frustration becomes the largest, as shown in Fig.~\ref{fig:fig2}(a), whose spin configuration is represented by Eq.~(\ref{eq:SkX}). 
Thus, the competing interactions in momentum space in the presence of $A$ turns out to be the microscopic origin of the S-SkX. 
In other words, it is important to take into account the contributions from the higher harmonics of $\bm{Q}_1$ and $\bm{Q}_2$ for $\zeta \simeq 1$. 
Furthermore, while increasing $A$, the stability region of the S-SkX becomes larger especially for $\zeta>1$, as shown in Figs.~\ref{fig:fig2}(b) and \ref{fig:fig2}(c), which means that the contributions from other $\bm{q}$ can be significant even when $\zeta$ is apart from 1 for relatively large $A$. 

In addition to the S-SkX, we find that the R-SkX phase for $\zeta \simeq 1$ by decreasing $H$ from the S-SkX phase. 
In the R-SkX, the skyrmion core is elongated so as to break the fourfold rotational symmetry of the square lattice, as shown in the lower panel of Fig.~\ref{fig:fig2}. 
The breaking of the fourfold rotational symmetry is found in the inequivalence between the $\bm{Q}'_1$ and $\bm{Q}'_2$ components of $\bm{m}_{\bm{q}}$ while keeping $\bm{m}_{\bm{Q}_1}=\bm{m}_{\bm{Q}_2}$ shown in Table~\ref{table:mp}. 
As this state exhibits a quantized skyrmion number of one, one expects a similar topological Hall effect to the S-SkX. 
In contrast to the S-SkX, the instability toward the R-SkX is enhanced for moderate $A$, as compared to the stability region for different $A$ in Fig.~\ref{fig:fig2}; large $A$ suppresses the instability toward the R-SkX.

\begin{table}[htb!]
\centering
\caption{
Nonzero components of $\bm{m}_{\bm{Q}_\eta}$ and $\bm{m}_{\bm{Q}'_\eta}$ ($\eta=1,2$) in each phase. 
\label{table:mp}}
\vspace{2mm}
\renewcommand{\arraystretch}{1.2}
\begin{tabular}{lccccccccccccccccccc}
\hline\hline
phase & $\bm{m}_{\bm{Q}'_1}$, $\bm{m}_{\bm{Q}'_2}$ ($\bm{Q}'_1 \parallel [100]$) & $\bm{m}_{\bm{Q}_1}$, $\bm{m}_{\bm{Q}_2}$ ($\bm{Q}_1 \parallel [110]$)\\ \hline
1$Q'$ VS & $m^{xy}_{\bm{Q}'_1}$, $m^{z}_{\bm{Q}'_1}$ & -- \\ 
1$Q'$ CS & $m^{xy}_{\bm{Q}'_1}$ & --\\ 
1$Q$ VS & --& $m^{xy}_{\bm{Q}_1}$, $m^{z}_{\bm{Q}_1}$\\ 
1$Q$ CS & --& $m^{xy}_{\bm{Q}_1}$\\  
\hline
R-SkX & $m^{xy}_{\bm{Q}'_1}, m^{z}_{\bm{Q}'_1}, m^{xy}_{\bm{Q}'_2}, m^{z}_{\bm{Q}'_2}$ &$m^{xy}_{\bm{Q}_1}=m^{xy}_{\bm{Q}_2}$, $m^{z}_{\bm{Q}_1}=m^{z}_{\bm{Q}_2}$\\ 
S-SkX &$m^{xy}_{\bm{Q}'_1}=m^{xy}_{\bm{Q}'_2}$, $m^{z}_{\bm{Q}'_1}=m^{z}_{\bm{Q}'_2}$ & $m^{xy}_{\bm{Q}_1}=m^{xy}_{\bm{Q}_2}$, $m^{z}_{\bm{Q}_1}=m^{z}_{\bm{Q}_2}$\\ 
\hline
2$Q$ I &$m^{xy}_{\bm{Q}'_1}, m^{z}_{\bm{Q}'_1}, m^{xy}_{\bm{Q}'_2}, m^{z}_{\bm{Q}'_2}$ &$m^{xy}_{\bm{Q}_1}=m^{xy}_{\bm{Q}_2}$, $m^{z}_{\bm{Q}_1}=m^{z}_{\bm{Q}_2}$ \\ 
2$Q$ II & $m^{z}_{\bm{Q}'_1}$ (or zero) & $m^{xy}_{\bm{Q}_1}=m^{xy}_{\bm{Q}_2}$\\ 
2$Q$ III & $m^{xy}_{\bm{Q}'_1}=m^{xy}_{\bm{Q}'_2}$ & $m^{xy}_{\bm{Q}_1}$, $m^{z}_{\bm{Q}_1}$,$m^{xy}_{\bm{Q}_2}$\\ 
2$Q$ IV & $m^{xy}_{\bm{Q}'_1}=m^{xy}_{\bm{Q}'_2}$, $m^{z}_{\bm{Q}'_1}=m^{z}_{\bm{Q}'_2}$ & $m^{xy}_{\bm{Q}_1}$, $m^{z}_{\bm{Q}_1}$,$m^{xy}_{\bm{Q}_2}$,$m^{z}_{\bm{Q}_2}$\\ 
2$Q$ V & $m^{xy}_{\bm{Q}'_1}=m^{xy}_{\bm{Q}'_2}$, $m^{z}_{\bm{Q}'_1}=m^{z}_{\bm{Q}'_2}$ & $m^{xy}_{\bm{Q}_1}$, $m^{z}_{\bm{Q}_2}$\\ 
2$Q$ VI &$m^{xy}_{\bm{Q}'_1}$, $m^{z}_{\bm{Q}'_1}$ & $m^{xy}_{\bm{Q}_1}=m^{xy}_{\bm{Q}_2}$\\ 
2$Q$ VII & $m^{xy}_{\bm{Q}'_1}, m^{xy}_{\bm{Q}'_2}$ & $m^{z}_{\bm{Q}_1}, m^{xy}_{\bm{Q}_2}$\\ 
2$Q$ VIII & $m^{xy}_{\bm{Q}'_1}, m^{z}_{\bm{Q}'_1}, m^{xy}_{\bm{Q}'_2}, m^{z}_{\bm{Q}'_2}$& $m^{xy}_{\bm{Q}_1}=m^{xy}_{\bm{Q}_2}$, $m^{z}_{\bm{Q}_1}=m^{z}_{\bm{Q}_2}$\\ 
\hline\hline
\end{tabular}
\end{table}
\begin{figure}[t!]
\begin{center}
\includegraphics[width=1.0 \hsize ]{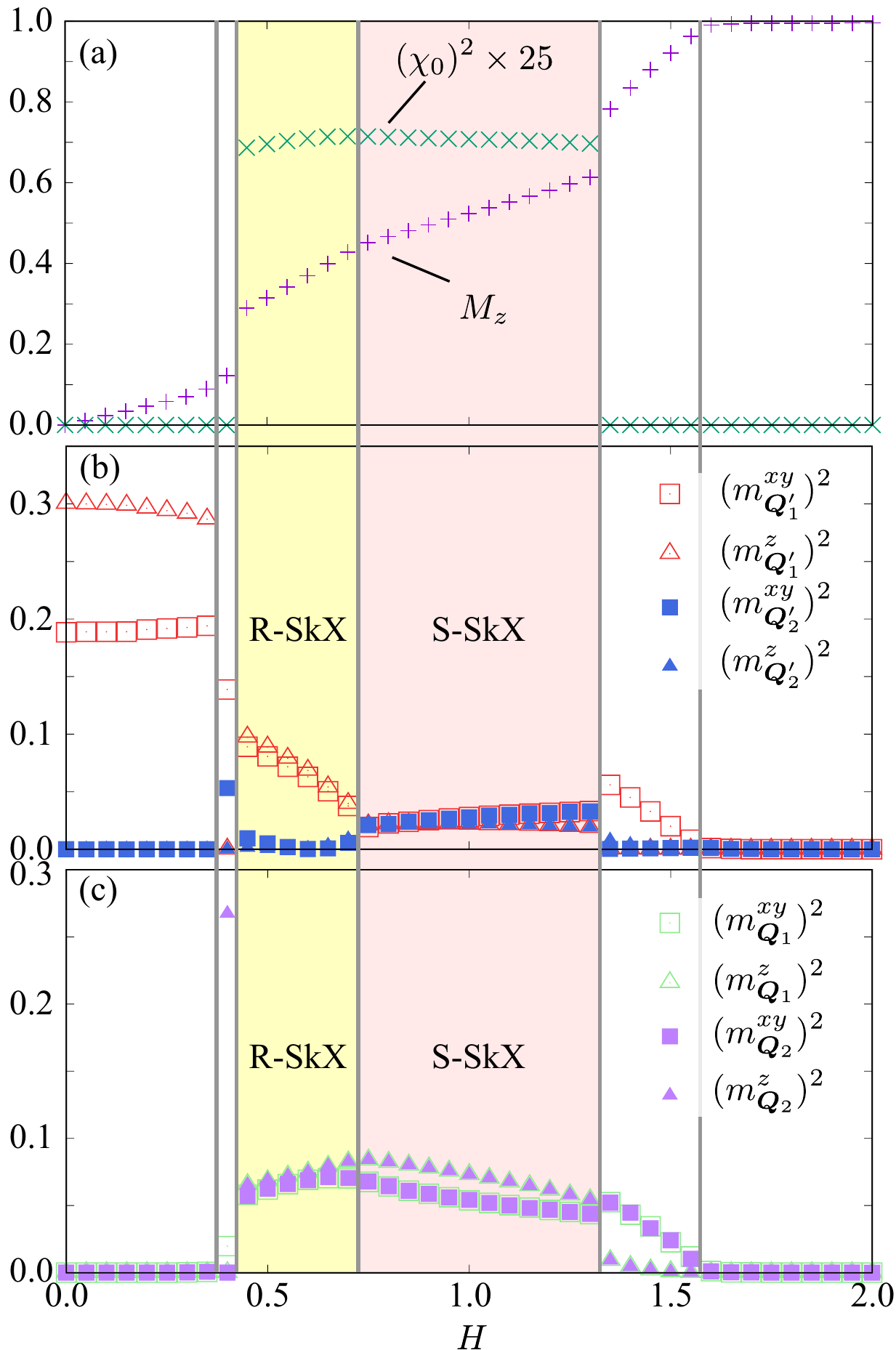} 
\caption{
\label{fig:fig3}
$H$ dependences of (a) $M_z$ and $(\chi_0)^2$, (b) $\bm{m}_{\bm{Q}'_\eta}$, and (c) $\bm{m}_{\bm{Q}_\eta}$ ($\eta=1,2$) at $A=0.2$ and $\zeta=1$. 
The vertical lines represent the phase boundaries in Fig.~\ref{fig:fig2}(b). 
}
\end{center}
\end{figure}

We show the $H$ dependence of the magnetization $M_z$ and the scalar chirality $\chi_0$ in Fig.~\ref{fig:fig3}(a), $\bm{m}_{\bm{Q}'_\eta}$ in Fig.~\ref{fig:fig3}(b), and $\bm{m}_{\bm{Q}_\eta}$ in Fig.~\ref{fig:fig3}(c) at $A=0.2$ and $\zeta=1$. 
Here, $\chi_0=(1/N)\sum_{i,\delta=\pm1} \bm{S}_i \cdot (\bm{S}_{i+\delta \hat{x}}\times \bm{S}_{i+\delta \hat{y}})$ where $\hat{x}$ ($\hat{y}$) is the unit vector in the $x$ ($y$) direction~\cite{Yi_PhysRevB.80.054416}. 
As shown in Fig.~\ref{fig:fig3}(a), the only two SkXs exhibit nonzero $\chi_0$, which are smoothly connected while changing $H$. 
$M_z$ also changes continuously, although its slope is slightly altered.
As discussed above, the S-SkX are characterized by nonzero $\bm{m}_{\bm{Q}_1}=\bm{m}_{\bm{Q}_2}$ and $\bm{m}_{\bm{Q}'_1}=\bm{m}_{\bm{Q}'_2}$, while the R-SkX has $\bm{m}_{\bm{Q}_1}=\bm{m}_{\bm{Q}_2}$ but $\bm{m}_{\bm{Q}'_1} \neq \bm{m}_{\bm{Q}'_2}$ reflecting the breaking of the fourfold rotational symmetry [Figs.~\ref{fig:fig3}(b) and \ref{fig:fig3}(c)], although their change in the transition seems to be continuous. 
The result indicates that a continuous transformation between the R-SkX and S-SkX is possible while keeping the topological property. 

The results of the phase diagrams in Fig.~\ref{fig:fig2} show that the S-SkX can appear for both $\zeta>1$ and $\zeta<1$. 
Especially, the situation for $\zeta<1$ indicates that $J_{\bm{Q}_1}$ and $J_{\bm{Q}_2}$ are not necessarily to be maxima in $J_{\bm{q}}$ for the stabilization of the S-SkX when the contributions from $J_{\bm{Q}'_1}$ and $J_{\bm{Q}'_2}$ are comparable to those from $J_{\bm{Q}_1}$ and $J_{\bm{Q}_2}$ for large $A$. 
Meanwhile, the R-SkX is stabilized for $\zeta>1$ in the present study, although it was shown that an additional bond-dependent anisotropy can stabilize the R-SkX even in the region for $\zeta<1$~\cite{hayami2022}. 

In addition to the two SkXs, we find various types of 2$Q$ states in Fig.~\ref{fig:fig2}, where nonzero $\bm{m}_{\bm{Q}_\eta}$ and $\bm{m}_{\bm{Q}'_\eta}$ are summarized in Table~\ref{table:mp}. 
Among them, the 2$Q$ I and 2$Q$ IV states, which are stabilized below the S-SkX phase in Fig.~\ref{fig:fig2}, exhibit a characteristic real-space spin configuration; the core denoted at $S_i^z\simeq -1$ forms the square lattice similar to the S-SkX, while there is no topological charge. 
Indeed, both states show local scalar chirality without global one like the meron-antimeron crystal~\cite{Lin_PhysRevB.91.224407,yu2018transformation,hayami2018multiple,kurumaji2019skyrmion,Hayami_PhysRevB.104.094425,Kato_PhysRevB.104.224405}. 
In this way, the competing interactions in momentum space give rise to rich multiple-$Q$ spin textures. 

\section{Summary}
\label{sec:Summary}
To summarize, we have investigated the stability of the SkX on the square lattice with single-ion anisotropy by taking into account the effect of the momentum-space frustration. 
By constructing the low-temperature phase diagrams on the basis of the simulated annealing, we found an instability tendency toward the S-SkX and R-SkX. 
Especially, we found that the realization of the R-SkX has been achieved by considering the synergy between the competing interactions in momentum space and the easy-axis single-ion anisotropy. 
We also obtained a variety of multiple-$Q$ states while changing the model parameters systematically.
The recent experiments indicate the rich multiple-$Q$ states in centrosymmetric tetragonal magnets like GdRu$_2$Si$_2$~\cite{khanh2020nanometric,Yasui2020imaging,khanh2022zoology},  EuAl$_4$~\cite{Shang_PhysRevB.103.L020405,kaneko2021charge,Zhu2022,takagi2022square}, EuGa$_4$~\cite{zhang2021giant,Zhu2022}, EuGa$_2$Al$_2$~\cite{moya2021incommensurate}, and Mn$_{2-x}$Zn$_x$Sb~\cite{Nabi_PhysRevB.104.174419}. 
Our phase diagrams to cover the wide range of model parameters will be a reference to understand the microscopic origin of the experimental findings.

Finally, let us discuss how to observe the signature of the competing interactions in momentum space in experiments. 
As the momentum-space competing interactions are the consequence of the real-space ones, the magnetic susceptibility measurement in the high-temperature region is a useful way of evaluating the short-range spin interactions that are relevant to the degree of frustration in the system, as often carried out in the frustrated magnets. 
Meanwhile, when considering itinerant magnets with the long-ranged Ruderman-Kittel-Kasuya-Yosida interaction, the information in terms of the Fermi surface and the band structure is important owing to the nesting property in metals. 
Thus, experiments, such as the angle-resolved photoemission spectroscopy and the de Haas-van Alphen effect, would be useful to identify the frustration in momentum space. 
In addition, the direct evaluation of the momentum-space interaction based on the ab-initio calculations might be desired for a further quantitative discussion~\cite{Nomoto_PhysRevLett.125.117204}.

\begin{acknowledgments}
This research was supported by JSPS KAKENHI Grants Numbers JP19K03752, JP19H01834, JP21H01037, and by JST PRESTO (JPMJPR20L8).
Parts of the numerical calculations were performed in the supercomputing systems in ISSP, the University of Tokyo.
\end{acknowledgments}

\bibliographystyle{apsrev}
\bibliography{ref}

\end{document}